\documentclass[12pt]{article}
\usepackage{amsmath,amsthm}

\begin{document}
\begin{center}
{\Large{\bf 
Superspace interpretation of mass--dependent \\
\smallskip 
BRST symmetries in general gauge theories\footnote{
Talk given at XXI Encontro Nacional de Fisica de
Particulos e Campos, October~23~--~27, 2000, 
S{\~a}o Lourenco, MG, Brazil.}
}}
\\
\bigskip\bigskip
{\large{B. Geyer}}$^{\, a,}$\footnote{On leave from 
Universit\"at Leipzig, Naturwissenschaftlich--Theoretisches Zentrum
and Institut f\"ur Theoretische Physik,
 04109 Leipzig, Germany; e-mail: geyer@itp.uni-leipzig.de}
{\large{and D. M\"ulsch}}$^{\, b}$
\\
\smallskip
$^{(a)}$ {\it Instituto de Fisica, Universidade de S{\~a}o Paulo,
05315-970 S{\~a}o Paulo, SP, Brazil}
\\
\smallskip
$^{(b)}${\it Wissenschaftszentrum Leipzig e.V., Leipzig 04103, Germany}

\bigskip
\end{center}

\begin{abstract}

A superspace formulation is proposed for the 
$osp(1,2)$--covariant Lagrangian quantization of general massive gauge 
theories. Thereby, $osp(1,2)$ is considered as subalgebra of 
the superalgebra $sl(1,2)$ which is interpreted as conformal
algebra acting on 
two anticommuting coordinates. The mass--dependent (anti)BRST symmetries 
of the quantum action in the osp(1,2) superfield formalism are 
realized as translations associated by
mass--dependent special conformal transformations.
\end{abstract}



\bigskip

\noindent
{\large{\bf 1 Introduction and main results}}

\medskip

\noindent
Recently, the $Sp(2)$--covariant Lagrangian quantization of Batalin, 
Lavrov and Tyutin~\cite{10} has been extended to a formalism which is 
based on the orthosymplectic 
superalgebra $osp(1,2)$ \cite{6} and which can be applied to {\it massive} 
gauge theories. This is achieved by incorporating into the extended BRST 
transformations $m$--dependent terms in such a way that the $m$--extended 
(anti)BRST symmetry of the quantum action $W_m$ is preserved. 
In that approach $W_m$ is required to satisfy the
generalized quantum master equations of $m$--extended BRST symmetry 
$(a = 1,2)$ and of $Sp(2)$ symmetry $(\alpha = 0,\pm 1)$,
\begin{eqnarray}
\label{qme1}
\bar{\Delta}^a \exp\{(i/\hbar)W_m\}=0 
\quad\Longleftrightarrow\quad
\hbox{$\frac{1}{2}$} ( W_m, W_m )^a + V_m^a W_m
&=& i \hbar \Delta^a W_m,
\\ 
\label{qme2}
\bar{\Delta}_\alpha \exp\{(i/\hbar)W_m\}=0
\quad\Longleftrightarrow\quad
\hbox{$\frac{1}{2}$} \{ W_m, W_m \}_\alpha + V_\alpha W_m 
&=& i \hbar \Delta_\alpha W_m,
\end{eqnarray}
respectively, whose generating (second order) differential operators
\begin{eqnarray} 
\label{Delta1}
\bar{\Delta}_m^a := \Delta^a + (i/\hbar) V_m^a,
\qquad 
\bar{\Delta}_\alpha := \Delta_\alpha + (i/\hbar) V_\alpha,
\end{eqnarray}
(for explicit expressions see Sect. 3)
form a superalgebra isomorphic to $osp(1,2)$  \cite{15}:
\begin{alignat}{3}
\label{IV46}
\hspace{-.5cm}
[ \bar{\Delta}_\alpha, \bar{\Delta}_\beta ] &= 
\hbox{\large$\frac{i}{\hbar}$}
\epsilon_{\alpha\beta}^{~~~\!\gamma} \bar{\Delta}_\gamma, 
&\quad~
[ \bar{\Delta}_\alpha, \bar{\Delta}_m^a ] &= 
\hbox{\large$\frac{i}{\hbar}$}
\bar{\Delta}_m^b (\sigma_\alpha)_b^{~a}, 
&\quad~
\{ \bar{\Delta}_m^a, \bar{\Delta}_m^b \} &= - m^2
\hbox{\large$\frac{i}{\hbar}$}
(\sigma^\alpha)^{ab} \bar{\Delta}_\alpha,
\end{alignat}
where the matrices $\sigma_\alpha$ generate the 
(real) Lie algebra $sl(2)$ being isomorphic to $sp(2)$ and the
$Sp(2)$--indices are raised or lowered by the (antisymmetric) tensor
$\epsilon^{ab},\, \epsilon^{12} = 1$, and $\epsilon_{0+-} =1$.
As long as $m \neq 0$ the operators $\bar{\Delta}_m^a$ are neither nilpotent
nor do they anticommute among themselves.
This algebra (without the factors $i/\hbar$) independently also holds for
$(V_m^a, V_\alpha)$.

The incorporation of mass terms into the action of any general
gauge theory is necessary at least 
intermediately within the BPHZL--renormalization scheme \cite{BPHZL}
which -- being independent of any regularization -- appears to be the 
most attractive one in order to formulate the quantum master equations 
on the level of algebraic renormalization theory. In that scheme 
by using Zimmermann's normal product formalism the r.h.~sides of 
Eqs.~(\ref{qme1}) and (\ref{qme2}) can be given a well-defined meaning
such that also higher--loop anomalies can be properly computed \cite{7}. 
In the BPHZL--scheme for any massless field a 
regularizing mass $m = (s - 1) M$ is introduced in order to be able to
perform besides ultraviolet also infrared subtractions thereby avoiding
spurious infrared singularities in the limit $s \rightarrow 1$. By using 
such an infrared regularization -- without violating the extended BRST 
symmetries -- the $osp(1,2)$--superalgebra occurs necessarily.

Here, we report a superfield representation \cite{GM}
of our earlier work on the
$osp(1,2)$--covariant Lagrangian quantization which amounts to understand
also the geometrical meaning of the $m$--dependent part of the extended BRST
transformations. For that reason we consider 
$osp(1,2)$ as subsuperalgebra of the superalgebra $sl(1,2)$. This
algebra, being isomorphic to $osp(2,2)$, contains four bosonic generators 
$V_\alpha$ and $V$, which form the Lie algebra $sl(2) \oplus u(1)$, and 
four (nilpotent) fermionic generators $V_+^a$ and $V_-^a$. 
The (anti)commutation relations of the superalgebra $sl(1,2)$ are 
\cite{15}: 
\begin{alignat}{3}
\label{II3}
[ V, V_\alpha ] &= 0,
&\qquad
[ V_\alpha, V_\beta ] &= \epsilon_{\alpha\beta}^{~~~\!\gamma} V_\gamma, 
&\qquad
\{ V_\pm, V_\pm \} &= 0,
\\
[ V, V_\pm^a ] &= \pm V_m^a,
&\qquad 
[ V_\alpha, V_\pm^a] &= V_\pm^b (\sigma_\alpha)_b^{~a}, 
&\qquad
\{ V_+^a, V_-^b \} &= - (\sigma^\alpha)^{ab} V_\alpha - \epsilon^{ab} V, 
\nonumber
\end{alignat}
The eigenvalues of the generators $V_\alpha$, for $\alpha = 0$, define 
the ghost numbers, whereas the eigenvalues of the generator $V$ define the 
Weyl weights which in Ref.~\cite{10} were introduced as `new ghost
number'. The generators $V_+^a$ and $V_-^a$ have opposite new ghost numbers,
${\rm ngh}(V_\pm^a) = \pm 1$, respectively. But, introducing a mass 
$m$ which formally will be attributed also by a new ghost number,
${\rm ngh}(m) = 1$, they can be combined into two fermionic 
generators $V_m^a = V_+^a + \hbox{$\frac{1}{2}$} m^2 V_-^a$ of the 
superalgebra $osp(1,2)$, Eqs.~(\ref{IV46}). (The difference 
$V_m^a = V_+^a - \hbox{$\frac{1}{2}$} m^2 V_-^a$
leads to another $osp(1,2)$--algebra.)

The key observation allowing for a geometric interpretation of the
superalgebra $sl(1,2)$ consists in interpreting it --
due to Baulieu, Siegel and Zwiebach \cite{11} --
as the algebra generating conformal transformations in a 
(super)space of two anticommuting coordinates $\theta^a$.
Hence, the generators $iV_+^a, iV_-^a,
iV^{ab} = i (\sigma^\alpha)^{ab} V_\alpha$ and $-iV$ 
may be considered as generators of translations 
$P^a$, special conformal transformations $K^a$, symplectic rotations 
$M^{ab}$ and dilatations $D$, respectively, in that space. 
This leads to a `natural' geometric interpretation of the $osp(1,2)$ 
quantization of general massive gauge theories: \\
The $osp(1,2)$--covariant quantization rules 
are formulated in terms of superfields $\Phi^A(\theta)$ and
superantifields $\overline\Phi_A(\theta)$
on which the generators of the algebra $sl(1,2)$ act linearly. 
The invariance of $W_m(\Phi, \overline\Phi)$ under $m$--extended 
BRST-- and $Sp(2)$--transformations is required through
Eqs.~(\ref{qme1}) and (\ref{qme2}), where 
$V_m^a = V_+^a + \hbox{$\frac{1}{2}$} m^2 V_-^a$ correspond to translations 
combined with $m$--dependent special conformal transformations and
$V_\alpha$ corresponds to symplectic rotations. Furthermore, proper 
solutions $S_m(\Phi, \overline\Phi)$ of the classical master 
equations $\hbox{$\frac{1}{2}$} ( S_m, S_m )^a + V_m^a S_m = 0$ and 
$\{ S_m, S_m \}_\alpha + V_\alpha S_m = 0$ with vanishing new ghost number,
${\rm ngh}(S_m) = 0$, correspond to solutions being 
invariant under dilatations, generated by $V$. Therefore,
these solutions are invariant under $osp(1,2) \oplus u(1)$, where the 
additional $u(1)$ symmetry is related to the new ghost number conservation.
In Ref.~\cite{GM} also the problem of how to determine
the transformations of the gauge fields and the full set of the necessary
(anti)ghost and auxiliary fields under the superalgebra $sl(1,2)$ has been
solved both for irreducible and first--stage reducible theories with closed
algebra (these results will not be reproduced here).
Finally, it is proven that mass terms generally destroy gauge independence
in the $osp(1,2)$--approach.
However, this gauge dependence disappears in the limit $m = 0$, thus
showing that the $osp(1,2)$--approach allows for a well-defined  
consideration of the renormalization of general gauge theories 
within the field--antifield formalism.

\bigskip

\noindent
{\large{\bf 2\, Superspace representations of $sl(1,2)$}}

\medskip

\noindent
In the $osp(1,2)$--approach the space of fields $\phi^A$ and antifields 
$\bar\phi_A, 
\phi^*_{Aa}$ and sources $\eta_A$ together with their Grassmann parities 
is characterized by the following sets \cite{10,6}:
\begin{alignat*}{2}
\phi^A &= ( A^i,~ B^{\alpha_s| a_1 \cdots a_s},~  
C^{\alpha_s| a_0 \cdots a_s}, s = 0, \ldots L ), 
&\quad~
\epsilon(\phi^A) &\equiv \epsilon_A = 
( \epsilon_i, \epsilon_{\alpha_s}\!+\!s, \epsilon_{\alpha_s}\!+\!s\!+\!1)
\\
\bar{\phi}_A &= ( \bar{A}_i,~ \bar{B}_{\alpha_s| a_1 \cdots a_s},~ 
\bar{C}_{\alpha_s| a_0 \cdots a_s}, s = 0,\ldots L ),
&\quad~
\epsilon(\bar{\phi}_A) &= \epsilon_A,
\\
\phi^*_{A a} &= ( A^*_{i a}, B^*_{\alpha_s a| a_1 \cdots a_s},
C^*_{\alpha_s a| a_0 \cdots a_s} , s = 0, \ldots L ),
&\quad~
\epsilon(\phi^*_{A a}) &= \epsilon_A + 1,
\\
\eta_A &= (D_i,~ E_{\alpha_s| a_1 \cdots a_s},~ 
F_{\alpha_s| a_0 \cdots a_s}, s = 0,\ldots L),
&\quad~
\epsilon(\eta_A) &= \epsilon_A,
\end{alignat*}
respectively.
Here, the pyramids of auxiliary fields $B^{\alpha_s| a_1 \cdots a_s}$ and  
(anti)ghosts $C^{\alpha_s| a_0 \cdots a_s}$  are irreducible
$Sp(2)$--tensors of `spin' $j=s$ and $j=s + 1$, respectively, 
being completely {\it symmetric} with 
respect to the `internal' indices $a_i=1,2,\;(i=0,1,\ldots,s)$;
similarly for 
$\bar\phi_A, \phi^*_{Aa}$ and 
$\eta_A$. The `external' index $a=1,2$ on the
$Sp(2)$--spinors $\phi^*_{Aa}$ is independent. 

Now, we introduce the sets of superfields
$\Phi^A(\theta)$ and superantifields $\overline{\Phi}_A(\theta)$ 
having equal Grassmann parity, 
$\epsilon({\Phi}^A) = \epsilon(\overline{\Phi}_A)\equiv \epsilon_A$, 
opposite ghost number, $gh(\overline{\Phi}_A) = - gh(\Phi^A)$,
and the following expansion in terms of component fields,
\begin{align}
\label{III24}
\Phi^A(\theta) &= \phi^A + \pi^{A a} \theta_a - \lambda^A \theta^2,
\qquad
\frac{\delta}{\delta \Phi^A(\theta)} = \frac{\delta}{\delta \phi^A} \theta^2 -
\theta^a \frac{\delta}{\delta \pi^{A a}} -
\frac{\delta}{\delta \lambda^A}
\\
\label{III25}
\overline{\Phi}_A(\theta) &= \bar{\phi}_A - \theta^a \phi^*_{A a} -
\theta^2 \eta_A,
\qquad
\frac{\delta}{\delta \overline{\Phi}_A(\theta)} =
\theta^2 \frac{\delta}{\delta \bar{\phi}_A} +
\frac{\delta}{\delta \phi^*_{A a}} \theta_a -
\frac{\delta}{\delta \eta_A}.
\end{align} 
According to DeWitt's convention derivatives 
with respect to the fields act from the {\em right}. 
Here, additional auxiliary fields 
$\pi^{Aa}$ and $\lambda^A$ have been introduced (cf.~also Ref.~\cite{4}).

In terms of the superantifields the representation of the generators 
of $sl(1,2)$ by {\em linear} differential operators on the superspace reads
\begin{align}
\label{III31}
V_+^a &= \int d^2 \theta \,
\frac{\partial \overline{\Phi}_A(\theta)}{\partial \theta_a}
\frac{\delta}{\delta \overline{\Phi}_A(\theta)},
\\
\label{III32}
V_-^a &= \int d^2 \theta \, \Bigr\{
2 \theta^2 \frac{\partial \overline{\Phi}_A(\theta)}{\partial \theta_a} + 
\theta_b \overline{\Phi}_B(\theta) \bigr(
(\sigma^\alpha)^{ab} (\sigma_\alpha)^B_{~~\!A} -
\epsilon^{ab} {\bar\gamma}^B_A \bigr) \Bigr\}
\frac{\delta}{\delta \overline{\Phi}_A(\theta)},
\\
\label{III33}
V_\alpha &= \int d^2 \theta \, \Bigr\{
- \theta_a (\sigma_\alpha)^a_{~b}
\frac{\partial \overline{\Phi}_A(\theta)}{\partial \theta_b} + 
\overline{\Phi}_B(\theta) (\sigma_\alpha)^B_{~~\!A} \Bigr\}
\frac{\delta}{\delta \overline{\Phi}_A(\theta)},
\\
\label{III34}
V &= \int d^2 \theta \, \Bigr\{
\theta_a \frac{\partial \overline{\Phi}_A(\theta)}{\partial \theta_a} + 
\overline{\Phi}_B(\theta) {\bar\gamma}^B_A \Bigr\}
\frac{\delta}{\delta \overline{\Phi}_A(\theta)},
\end{align}
where $ (\sigma_\alpha)^B_{~~\!A}$ are irreducible $Sp(2)$--representations
of spin $j$ acting on the set of internal $Sp(2)$-indices, and 
${\bar\gamma}^B_A= \alpha(\overline{\Phi}_A) \delta^B_A$ is related to 
the Weyl weight $\alpha(\overline{\Phi}_A)$ of the antisuperfields 
${\overline\Phi}_A$, coinciding with their new ghost number and obeying
$\alpha(\overline{\Phi}_A) + \alpha(\Phi^A) = -2$ (for details see
\cite{GM}). 

Replacing in Eqs.~(\ref{III31})--(\ref{III34}) the superantifield
$\overline{\Phi}_A(\theta)$ by the superfield $\Phi^A(\theta)$, 
the left derivatives
$\delta_L/ \delta \overline{\Phi}_A(\theta)$ by the right--derivatives 
$\delta_R/ \delta \Phi^A(\theta)$, $\bar\gamma^B_A$ by $\gamma^B_A$,
and reversing the order of all the factors, then the corresponding 
{\em linear} right--representation $(U_\pm^a, U_\alpha, U)$
of the $sl(1,2)$--algebra on the superfields is obtained.
If the auxiliary fields $\pi^{Aa}$ and $\lambda^A$ finally are eliminated
from the action by integrating them out in the functinal integral,
cf.~Eq.~(\ref{V68}) below, then the {\em nonlinear} (anti)BRST 
transformations re-appear \cite{GM}.

\bigskip

\noindent
{\large{\bf 3\, $osp(1,2)$--covariant superfield quantization}}

\medskip

\noindent
In the superfield approach to the quantization of general gauge theories the
m--dependent quantum action $W_m(\Phi^A(\theta),\overline{\Phi}_A(\theta))$ 
cannot be required to be invariant under the whole superalgebra $sl(1,2)$.
Instead, it will be required to be invariant under only one of its two 
$osp(1,2)$--subalgebras. $W_m(\Phi^A(\theta),\overline{\Phi}_A(\theta))$
is assumed to be invariant under a mass-dependent combination of translations
 and special conformal transformations, symplectic rotations and, eventually,
 dilatations in $\theta^a$--space. The generators 
$\bar\Delta_m^a, \bar\Delta_\alpha$ and $\bar\Delta_m$, respectively,
of these symmetries will be introduced now explicitly.
 
The odd and even differential operators $\bar\Delta_m^a$
and $\bar\Delta_\alpha$, Eqs.~(\ref{Delta1}), respectively, are given 
on the space of superfields $\Phi^A(\theta)$ and superantifields 
$\overline{\Phi}_A(\theta)$ as follows:
\begin{align*}
\Delta^a &= \int d^2 \theta 
\frac{\partial^2 \delta_L}{\partial \theta^2 \delta \Phi^A(\theta)} \,
\theta^a \frac{\delta}{\delta \overline{\Phi}_A(\theta)} = 
(-1)^{\epsilon_A} \frac{\delta_L}{ \delta \phi^A} \, 
\frac{\delta}{ \delta \phi_{A a}^*},
\qquad
V_m^a \equiv V_+^a + \hbox{$\frac{1}{2}$} m^2 V_-^a,
\\
\Delta_\alpha &= (-1)^{\epsilon_A + 1} \int d^2 \theta \,  
\theta^2 (\sigma_\alpha)_B^{~~\!A} \frac{\partial^2 \delta_L}
{\partial \theta^2 \delta \Phi^A(\theta)} \,
\frac{\delta}{\delta \overline{\Phi}_B(\theta)} =
(-1)^{\epsilon_A} (\sigma_\alpha)_B^{~~\!A} 
\frac{\delta_L}{ \delta \phi^A} \frac{\delta}{\delta \eta_B},
\end{align*}
with the translation operators $V_+^a$ and the
special conformal operators $V_-^a$ given by Eqs.~(\ref{III31})
and (\ref{III32}), and the operators of symplectic rotations $V_\alpha$ 
given by Eq.~(\ref{III33}), respectively. The (second--order) 
differential operators $\Delta^a$ and $\Delta_\alpha$
are associated by two {\em odd} superantibrackets $(F,G)^a$  
and by three {\em even} superbrackets $\{F,G\}_\alpha$, respectively,
\begin{align*}
( F,G )^a &= (-1)^{\epsilon_A} \int d^2 \theta \, \Bigr\{ 
\frac{\partial^2 \delta F}
{\partial \theta^2 \delta \Phi^A(\theta)} \theta^a 
\frac{\delta G}{\delta \overline{\Phi}_A(\theta)} - 
(-1)^{(\epsilon(F) + 1) (\epsilon(G) + 1)} (F \leftrightarrow G) \Bigr\},
\\
\{ F,G \}_\alpha &= - \int d^2 \theta \, \Bigr\{ 
\theta^2 \frac{\partial^2 \delta F}
{\partial \theta^2 \delta \Phi^A(\theta)} \,
\frac{\delta G}{\delta \overline{\Phi}_B(\theta)} (\sigma_\alpha)_B^{~~\!A} +
(-1)^{\epsilon(F) \epsilon(G)} (F \leftrightarrow G) \Bigr\}. 
\end{align*}

The quantum action $W_m (\Phi^A(\theta), \overline{\Phi}_A(\theta))$ 
is required to obey the 
$m$--extended generalized quantum master equations (\ref{qme1})
ensuring (anti)BRST invariance, and the generating equations (\ref{qme2})
ensuring $Sp(2)$--invariance.
The solution of these equations is sought of as a power series in Planck's
constant $\hbar$, 
$W_m = S_m + \sum_{n = 1}^\infty \hbar^n W_{m}^{(n)}$,
obeying the requirements of  nondegeneracy of $S_m$ and the correctness 
of the classical limit, i.e.,~that $S_m$ coincides with the classical 
action $S_{\rm cl}(A)$ if the superantifields are put equal to zero
(and the auxiliary fields $\pi^{Aa}$ and $\lambda^A$ are integrated out).
According to the definition of the superantifields the action $W_m$ 
depends on $\eta_A$ only linearly.

The gauge fixed quantum action 
$W_{m, {\rm ext}}(\Phi^A(\theta), \overline{\Phi}_A(\theta))$ is introduced
according to 
\begin{equation}
\label{IV50}
{\rm exp}\{ (i/ \hbar) W_{m, {\rm ext}} \} = 
\hat{U}_m(F) \,{\rm exp}\{ (i/ \hbar) W_m \}, 
\end{equation}
where the operator $\hat{U}_m(F)$ has to be choosen as \cite{6}
\begin{eqnarray}
\label{gauge}
\hat{U}_m(F) = {\rm exp}\{(\hbar/ i) \hat{T}_m(F)\}
\quad
{\rm with}
\quad
\hat{T}_m(F) = \hbox{$\frac{1}{2}$} \epsilon_{ab} 
\{ \bar{\Delta}_m^b, [ \bar{\Delta}_m^a, F ] \} + (i/ \hbar)^2 m^2 F,
\end{eqnarray}
$F(\Phi^A(\theta))$ being a $Sp(2)$--symmetric
 bosonic gauge fixing functional with vanishing ghost number. 
Restricting $W_m$ to the subspace of {\em admissible} actions
satisfying the requirement
$\int d\theta \theta^2 \left\{ \delta W_m/\delta {\overline\Phi}_A 
- \Phi^A\right\} = 0$,  
then
the gauge fixed quantum action $W_{m, {\rm ext}}$ also obeys the 
quantum master equations (\ref{qme1}) and (\ref{qme2}).

Let us now introduce a further differential operator
$\bar{\Delta}_m = \Delta + (i/\hbar) V_m$
according to 
\begin{eqnarray}
\Delta = (-1)^{\epsilon_A + 1} \int d^2 \theta \,  
\theta^2 \gamma_B^A \frac{\partial^2 \delta_L}
{\partial \theta^2\delta \Phi^A(\theta)} 
\frac{\delta}{\delta \overline{\Phi}_B(\theta)} = 
(-1)^{\epsilon_A}  \gamma_B^A \frac{\delta_L}{ \delta \phi^A} 
\frac{\delta}{\delta \eta_B},
\quad
V_m \equiv V + m \frac{\partial}{\partial m},
\nonumber
\end{eqnarray}
with the dilatation operator $V$ given by Eq.~(\ref{III34}). The 
differential operator $\Delta$ is associated by the following 
expression (being not a new bracket since $ \gamma_B^A$ is diagonal)
\begin{equation}
\label{IV56}
\{ F,G \} = - \int d^2 \theta \, \Bigr\{ 
\theta^2 \gamma_B^A \frac{\partial^2 \delta F}
{\partial \theta^2 \delta \Phi^A(\theta)} \,
\frac{\delta G}{\delta \overline{\Phi}_B(\theta)} +
(-1)^{\epsilon(F) \epsilon(G)} (F \leftrightarrow G) \Bigr\}.
\nonumber
\end{equation}
The additional operator $\bar{\Delta}_m$ together with 
$\bar{\Delta}_m^a$ and $\bar{\Delta}_\alpha$ forms a 
superalgebra being isomorphic to $osp(1,2) \oplus u(1)$ where,
in addition to the (anti)commutation relations (\ref{IV46}), 
the following commutation relations hold true
(analogously for ($V^a_m, V_\alpha, V_m$)):
\begin{alignat}{3}
\label{Delta_m}
[ \bar{\Delta}_m, \bar{\Delta}_m ] &= 0,
&\qquad
[ \bar{\Delta}_m, \bar{\Delta}_\alpha ] &= 0,
&\qquad
[ \bar{\Delta}_m, \bar{\Delta}_m^a ] &= 
\hbox{\large$\frac{i}{\hbar}$}\bar{\Delta}_m^a.
\end{alignat}

Let us now assume that solutions $W_m$ of the quantum master equations 
(\ref{qme1}) and (\ref{qme2}) can be constructed obeying new ghost number
conservation being expressed by the equation: 
\begin{equation}
\label{IV55}
\bar{\Delta}_m\, {\rm exp}\{ (i/ \hbar) W_m \} = 0
\qquad
\Longleftrightarrow 
\qquad
\hbox{$\frac{1}{2}$} \{ W_m, W_m \} + V_m W_m = i \hbar \Delta_m W_m.
\end{equation}
However, it is already well-known that the new ghost number is conserved 
only in the limit $\hbar \rightarrow 0$. In addition, the new ghost number 
conservation is broken also through gauge fixing \cite{GM}. 
Therefore, Eq.~(\ref{IV55})
should be required only for the tree approximation of the quantum action;
eventually, it could hold if no radiation corrections occure.

\bigskip

\noindent
{\large{\bf 4\, Generating functionals and gauge (in)dependence}}

\medskip

\noindent
The vacuum functional $Z_m(0)$ in the super(anti)field approach is defined as
\begin{equation}
\label{V68}
Z_m(0) = \int d \Phi^A(\theta) \, d \overline{\Phi}_A(\theta) \,
\rho(\overline{\Phi}_A(\theta))
\exp\{ (i/ \hbar) ( W_m - S_{m, F} + S_{m, X}) \},
\end{equation}
with
\begin{eqnarray*}
S_{m, X} &=&  \int d^2 \theta \, \overline{\Phi}_A(\theta) \Phi^A(\theta)
+ m^2 \int d^2 \theta \, \theta^2 
\overline{\Phi}_A(\theta) \gamma^A_B \Phi^B(\theta),
\\
S_{m, F} &=&
 \int d^2 \theta \Bigl\{
\frac{\delta F}{\delta \Phi^A(\theta)} 
\frac{\partial^2 \Phi^A(\theta)}{\partial \theta^2} + 
\hbox{\large $\frac{1}{2}$} \epsilon_{ab} \int d^2 \bar{\theta} \,
\frac{\partial \Phi^A(\theta)}{\partial \theta_a}
\frac{\delta^2 F}{\delta \Phi^A(\theta) \delta \Phi^B(\bar{\theta)}}
\frac{\partial \Phi^B(\bar{\theta})}{\partial \bar{\theta}_b} \Bigr\}
\nonumber\\
&&+ \hbox{\large $\frac{1}{2}$} m^2 
\int d^2 \theta \, \theta^2 
\frac{\partial^2 \delta F}{\partial \theta^2 \delta \Phi^A(\theta)}
\gamma^A_B \Phi^B(\theta),
\nonumber
\end{eqnarray*}
and the measure given by
$\rho(\overline{\Phi}^A)=
\delta\left(\int d^2\theta \overline{\Phi}(\theta)\right)
= \delta(\eta_A)$.

The integrand in (\ref{V68}) is invariant under the 
following global transformations:
\begin{align}
\label{V69}
\delta \Phi^A(\theta) &= \Phi^A(\theta) U_m^a \mu_a,
\qquad
\delta \overline{\Phi}_A(\theta) = \mu_a V_m^a \overline{\Phi}_A(\theta) +
\mu_a ( W_m, \overline{\Phi}_A(\theta) )^a
\\
\label{V70}
\delta \Phi^A(\theta) &= \Phi^A(\theta) U_\alpha \mu^\alpha,
\qquad
\delta \overline{\Phi}_A(\theta) = \mu^\alpha V_\alpha \overline{\Phi}_A(\theta) +
\mu^\alpha \{ W_m, \overline{\Phi}_A(\theta) \}_\alpha,
\end{align}
where $\mu_a, \epsilon(\mu_a) = 1$, and $\mu^\alpha, 
\epsilon(\mu^\alpha) = 0$, are constant anticommuting resp.~commuting
parameters. 
These transformations realize the $m$--extended 
(anti)BRST-- and $Sp(2)$--symmetry, respectively, in the superfield
approach to $osp(1,2)$--covariant quantization.

Let us now change the gauge fixing functional in (\ref{V68}) according to
$F \rightarrow F + \delta F$ followed by the transformations 
(\ref{V69}) with the choice
$ 
\mu_a = - (i/\hbar) \hbox{$\frac{1}{2}$} \epsilon_{ab} (\delta F) U_m^b.
$ 
This leads to
\begin{equation*}
S_{m, F} \rightarrow S_{m, F} + 
\left[\hbox{$\frac{1}{2}$} \epsilon_{ab} (\delta F) U_m^b U_m^a
+ m^2 \delta F\right]+(\hbar/i) \mu_a U_m^a  = 
S_{m, F} + m^2 \delta F.
\end{equation*}
Thus we observe that the mass term $m^2 F$ in Eq.~(\ref{gauge}) violates 
the independence of $Z_m(0)$ on the choice of the gauge. Unfortunaley, 
that unwanted term can not be compensated by any further change
of variables, thus showing that it breaks gauge independence of the
$S$--matrix. However, gauge independence is restored in the limit
$m \rightarrow 0$, i.e., $s \rightarrow 1$, which has to be taken
after having carried out all the ultraviolet and infrared subtractions.

One of the virtues of the quantization scheme presented here is
that, first, during the process of renormalization the $osp(1,2)$--symmetry
of the theory is maintained and, second, this enlarged symmetry --
in comparision with the usual field-antifield formalism -- allows for
a much easier, algebraic proof of possible absence of anomalies.
This formalism has been succesfully applied to the instanton sector
of QCD \cite{M} and to the quantization of Yang-Mills theories
in a generic background configuration \cite{GM0}.

\bigskip

\noindent
{\large{\bf Acknowledgement}}

\medskip

\noindent
One of the authors (B.G.) thankful acknowledges financial support
by German-Brazil exchange program of FAPESP and DAAD
during his stay at Institute of Physics of S{\~a}o Paulo University.

\end{document}